\documentclass[aps,prd,preprint,groupedaddress]{revtex4-2}
\usepackage{graphicx}
\usepackage{amsmath}
\usepackage{todonotes}
\newcommand \beq{\begin{eqnarray}}
\newcommand \eeq{\end{eqnarray}} 
\def\x{{\boldsymbol x}}

\def\b{{\boldsymbol b}}

\def\r{{\boldsymbol r}}

\def\s{{\boldsymbol s}}

\newcommand{\rmd}{{\rm d}}

\newcommand{\nn}{\nonumber\\ }
\begin{document}
\title{Phenomenological study of quarkonium suppression and the impact of the energy gap between singlets and octets}
\author{Jean-Paul Blaizot}
\affiliation{Institut de  Physique Th\'eorique,  Universit\'e Paris Saclay, 
        CEA, CNRS, 
	F-91191 Gif-sur-Yvette, France}
\author{Miguel \'{A}ngel Escobedo}
\affiliation{Instituto Galego de F\'{i}sica de Altas Enerx\'{i}as (IGFAE), Universidade
de Santiago de Compostela. E-15782, Galicia, Spain}
\date{\today}
\begin{abstract}
The study of heavy quarkonium suppression in heavy-ion collisions represents an important source of information about the properties of the quark-gluon plasma produced in such collisions. In a previous paper, we have considered how to model the evolution of a quarkonium in such a way that the solution of the resulting equations evolves toward the correct thermal equilibrium distribution for an homogeneous and static medium. We found that it is crucial to take into account the energy gap between singlet and octet configurations when the temperature is not much greater than this energy gap. In this manuscript, we explore in more detail the phenomenological consequences of this observation in the more realistic situation of an expanding system. We consider two different scenarios, based on the same approximation scheme, but on different choices of parameters. In the first case, we rely on a Hard Thermal Loop approximation, while the second case is  based on a recent determination of the static potential in lattice QCD. In both cases, we compute the decay width and the nuclear modification factor, both taking the energy gap into account and ignoring it. We find that the impact on the predictions is as large in the expanding medium as it is in the static case. Our conclusion is that this energy gap should be taken into account in phenomenological studies.
\end{abstract}
\maketitle
\section{Introduction}

 Several physical phenomena are commonly invoked to explain the production of quarkonia in ultra-relativistic heavy ion collisions that are presently intensively studied at LHC \cite{Sirunyan:2018nsz,Acharya:2020kls,Aaboud:2018quy}. Aside from the initial suggestion of the color screening of the binding potential by the quark-gluon plasma \cite{Matsui:1986dk}, 
collisions between the plasma constituents and the quarkonia could also lead to a suppression of the production rate. Various ways can be used to take these collisions into account. In the regime in which potential models are valid, it has been shown that collisions induce  an imaginary part in the potential \cite{Laine:2006ns,Beraudo:2007ky} (somewhat analogous to the imaginary potential of the nuclear optical potential model). In the effective theory picture where the interactions of quarkonia  with the plasma are dominated by color dipolar interactions, collisions induce singlet to octet transitions \cite{Brambilla:2008cx}. In addition to these ``suppression'' mechanisms, there is also the possibility, when the number of heavy quarks created in a heavy-ion collision is large enough, that uncorrelated heavy quarks  recombine to form a bound state inside  the medium. This process, usually called recombination \cite{BraunMunzinger:2000px,Thews:2000rj}, appears to be an important contribution to charmonium production at present collider energies.

The construction of a robust formalism that takes into account consistently  all three mechanisms is a challenging task.  In recent years, it has been found useful to consider the various approaches from the perspective of open quantum systems, the ``system'' being the quarkonia, interacting with the quark-gluon plasma considered as ``environment''  \cite{Borghini:2011ms,Akamatsu:2014qsa,Akamatsu:2018xim,Blaizot:2015hya,Brambilla:2016wgg,Brambilla:2017zei,Kajimoto:2017rel,Katz:2015qja,Yao:2018nmy,Yao:2018sgn,Blaizot:2017ypk,Blaizot:2018oev,Yao:2020eqy}. Recent reviews can be found in \cite{Akamatsu:2020ypb,Yao:2021lus}. This formalism allows us to compute the evolution of bound states in a medium, taking into account quantum mechanical effects. In this way, it has been  understood that quantum coherence \footnote{By this we mean that non diagonal elements of the density matrix play a role, not only the diagonal ones that are kept in elementary transport approaches.} is essential at high temperatures when the binding energy is of the order of the decay width. At the same time, semi-classical equations can be derived rigorously from the open quantum system framework in special limits \cite{Blaizot:2015hya,Blaizot:2017ypk,Blaizot:2018oev,Yao:2018nmy,Yao:2020eqy}. This has improved the understanding of the validity region of these semi-classical approaches. The common assumption in most, if not all, approaches is that the medium is moderately affected by the presence of  quarkonia. Various types of effective theories can then be obtained for the heavy quark dynamics, with further approximations becoming available, depending on the energy or time scales. 

In the limit in which the temperature, $T$, is much bigger than the binding energy, $E$, the interaction between a quarkonium and the medium is Markovian. In this case, the dynamics is governed by  a Gorini Kossakowski Sudarshan Lindblad (GKSL) equation \cite{Gorini:1975nb,Lindblad:1975ef}. Phenomenological predictions of quarkonium in heavy-ion collisions using the GKSL equation in the $T\gg E$ limit have been discussed in \cite{Brambilla:2016wgg,Brambilla:2017zei,Brambilla:2020qwo}. However, as was understood in \cite{Blaizot:2018oev}, this limit leads to a maximization of the entropy without taking into account energy conservation constraints. In other words, this approximation does not lead to the correct thermalization. In \cite{Blaizot:2018oev}, we showed that, when $T\sim E$, the decay width is modified to include a dependence on the binding energy. With a slight abuse of language, we can say that the imaginary part of the potential depends on the energy. This dependence is a key element which guarantees that the evolution equation complies with the fluctuation-dissipation theorem. Phenomenologically, it is important to take this dependence into account when the temperature is not much larger than the binding energy, as was shown numerically in \cite{Yao:2017fuc}. 

The aim of this manuscript is to emphasize the importance of the energy dependence of the decay width, not only in the case of a static medium, but also in a more realistic case of an expanding medium. We do this by computing the nuclear modification factor, $R_{AA}$, in different scenarios. For each scenario, we compare the results obtained with and without the  energy dependence of the imaginary part of the potential. We observe that, depending on the centrality of the collision, the productions of $\Upsilon(1S)$ and $\Upsilon(2S)$ are significantly less suppressed when we take into account the energy dependence. We emphasize that our purpose here is just to obtain an estimate of the effect, and show that it is indeed quite significant. We shall not go into a detailed analysis of experimental data. 

The manuscript is organized as follows. In section \ref{sec:model}, we review the model that we developed in \cite{Blaizot:2018oev} and we discuss the imaginary part of the potential and its energy dependence. In section \ref{sec:pheno}, we use this model to obtain phenomenological predictions of $R_{AA}$ using two different scenarios, one based on (resummed) perturbation theory,  and another scenario in which we use lattice QCD inputs. Finally, we present our conclusions in section \ref{sec:conclu}.

\section{Discussion of the model}
\label{sec:model} 
The model that we consider in the present paper is an extension of that introduced in the last section of \cite{Blaizot:2018oev}. Let us recall briefly its origin. One starts from the evolution equation for the reduced density matrix ${\cal D}_Q$ of a pair of heavy quarks in a quark-gluon plasma, the interaction of the heavy quarks with the plasma, supposed to be weak,  being limited to a one-gluon-exchange. The plasma is assumed to be in thermal equilibrium at a temperature $T$.  This equation reads \cite{Blaizot:2018oev}
\beq
\label{eqrhoAt0c2b2b2}
&&\!\!\!\!\!\!\!\!\!\! \frac{\rmd {\cal D}_Q}{\rmd t}+i[H_Q,{\cal D}_Q(t)]=\nn
&&-g^2 \int_{\x\x'}
 \int_{0}^{t-t_0} \rmd \tau \,[n^A_\x ,U_Q(\tau)n^A_{\x'}U_Q^\dagger(\tau) {\cal D}_Q(t)]  \, \Delta^>(\tau;\x-\x'))\nn
&&-g^2\int_{\x\x'}\int_{0}^{t-t_0} \rmd \tau \, [{\cal D}_Q(t)U_Q(\tau) n^A_{\x'}U_Q^\dagger(\tau), n^A_\x]  \,\Delta^<(\tau;\x-\x').
\eeq
Here, $H_Q$ is the Hamiltonian that governs the dynamics of the heavy quark pair in vacuum, with the corresponding evolution operator given by $U_Q(t)={\rm e}^{-iH_Qt}$.  $n^A_\x$ is the temporal component of the color current, whose explicit expression can be found in \cite{Blaizot:2017ypk}. The interaction between the heavy quarks and the surrounding medium is captured by the correlators of thermal fluctuations of the gluon fields $a_0^A$, 
\beq\label{correlators}
&&{\rm Tr}_{\rm pl}\left[a_0^A(t,{\bf x})a_0^B(t',{\bf y}){\cal D}_{\rm pl}\right]=\delta^{AB}\Delta^>(t-t',{\bf x}-{\bf y}),\nn
&&{\rm Tr}_{\rm pl}\left[a_0^B(t',{\bf y})a_0^A(t,{\bf x}){\cal D}_{\rm pl}\right]=\delta^{AB}\Delta^<(t-t',{\bf x}-{\bf y}),
\eeq
where ${\cal D}_{\rm pl}$ is the density matrix of the plasma, and $A,B$ are color indices of the adjoint representation. We ignore here the color magnetic interactions, which are subleading for heavy quarks. 

As was shown in detail in \cite{Blaizot:2018oev} this general  equation can be simplified through a number of steps:
\begin{itemize}
\item A part of the right-hand side  of Eq.~(\ref{eqrhoAt0c2b2b2}), which is hermitian,  can be absorbed in a redefinition of $H_Q$. This can be done by adding to both sides of Eq.~(\ref{eqrhoAt0c2b2b2})
\begin{equation}
\frac{g^2}{2} \int_{\x\x'}
 \int_{0}^{t-t_0} \rmd \tau \,[n^A_\x n^A_{\x'},{\cal D}_Q(t)]  \, (\Delta^>(\tau;\x-\x')-\Delta^<(\tau;\x-\x'))\,.
\end{equation}
In the left-hand side, this is considered as a correction to the real part of the potential, while in the right-hand side, it remains as a compensating correction.  From now on,  $H_Q$  denotes the heavy quark hamiltonian after this redefinition.
\item  By projecting Eq.~(\ref{eqrhoAt0c2b2b2}) on the eigenstates of  $H_Q$, with $H_Q|i\rangle=E_i|i\rangle$, one  gets
\begin{equation}
\frac{d{\cal D}_{ij}}{dt}+iE_{ij}{\cal D}_{ij}={\cal L}_{ij,kl}{\cal D}_{kl}\,,
\end{equation}
where ${\cal D}_{ij}=\langle i|{\cal D}_Q|j\rangle$, $E_{ij}=E_i-E_j$ and ${\cal L}_{ij,kl}{\cal D}_{kl}$ is a rewriting of the right-hand side  of Eq.~(\ref{eqrhoAt0c2b2b2}) from which the hermitian contribution to $H_Q$ has been subtracted.  
\item When the typical time between collisions of the quarkonium with plasma constituents is much bigger than the inverse of the binding energy, which we shall assume to be the case,  ${\cal L}_{ij,kl}$ can be treated as a perturbation. The multiple-scale analysis described  in appendix B of \cite{Blaizot:2018oev} provides a way to handle the secular terms which appear in the perturbative expansion. Assuming for simplicity that there are no degenerate states, we obtain then an equation for the diagonal part of the density matrix, namely for the probabilities ${p}_i=D_{ii}$
\begin{equation}
\frac{dp_i}{dt}={\cal L}_{ii,kk}p_k\,.
\label{eq:multi}
\end{equation}
\item A final simplification arises from the analysis of the color degrees of freedom. We are interested in the probability to find the quarkonium  in a  singlet state at the end of the evolution. By emitting or absorbing gluons, a color singlet can decay into an octet, and vice versa a singlet can emerge from an octet. We shall restrict ourselves to the dilute limit where there are  only a few heavy quarks in the medium. This situation is appropriate for bottomonium production at the LHC. The study of charmonium would  concern the dense limit with a large number of heavy quarks present in the medium. This is a much more difficult problem, which has only been addressed so far in special limits \cite{Blaizot:2017ypk} (see also \cite{Blaizot:2015hya} for the analogous abelian case), or using the  Boltzmann equation \cite{Yao:2018nmy,Yao:2018sgn}. In the case of bottomonium, we have checked numerically that the contribution from octet decays is tiny. The reason is that octets are unbound and may occupy all the volume of the medium,  while a bound color singlet occupies a small volume. Therefore, in the dilute limit, the rate of the octet to singlet transitions decreases as the volume of the medium increases. Besides, this transition is also suppressed in the large $N_c$ limit \footnote{This observation is also valid beyond perturbation theory \cite{Escobedo:2020tuc}}. Finally, we arrive at a simple rate equation  
\begin{equation}
\frac{dp_{\rm s}}{dt}=-\Gamma_{\rm s} \,p_{\rm s}\,,
\label{eq:modfin}
\end{equation}
for  the probability $p_{\rm s}$ of finding the quarkonium in a singlet state. In the next subsection we analyse the properties of the decay rate $\Gamma_{\rm s}$.
\end{itemize} 
\subsection{Properties of $\Gamma_{\rm s}$.}
\begin{figure}
\includegraphics[scale=1]{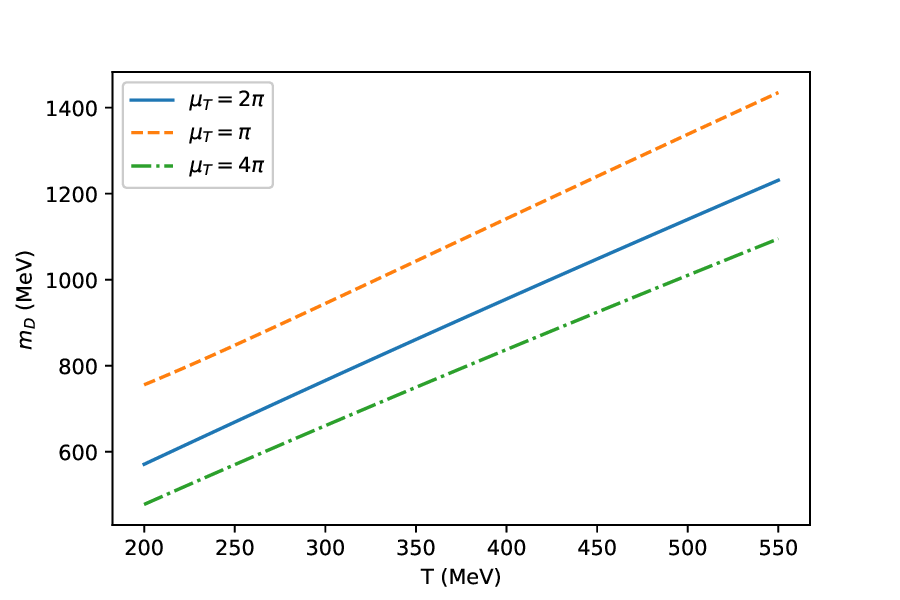}
\caption{Perturbative Debye mass as a function of the temperature, estimated from Eq.~(\ref{eq:md}) with the running coupling constant $\alpha_s$ evaluated at different values of the scale $\mu_T$.}
\label{fig:md}
\end{figure}

The decay width $\Gamma_{\rm s}$ is given by the following formula \cite{Blaizot:2018oev}
\begin{widetext}
\begin{equation}
\Gamma_{s}=8\pi\alpha_sC_{F}\int_{{\bf p}}{\rm e}^{-\frac{E_{{\bf p}}^{{\rm o}}-E^{{\rm s}}}{T}}\int_{{\bf q}}\Delta^{>}(E_{{\bf p}}^{{\rm o}}-E^{{\rm s}},{\bf q})|\langle{\rm s}|\sin\left(\frac{{\bf q}\cdot\hat{{\bf r}}}{2}\right)|{\rm o},{\bf p}\rangle|^{2}\,,
\label{eq:width}
\end{equation}
\end{widetext}
where the operator $\sin\left(\frac{{\bf q}\cdot\hat{{\bf r}}}{2}\right)$, with $\hat\r$ an operator acting on the heavy quark relative coordinates, describes the interaction of the quarkonium with the gluons of the plasma, after the center of mass of the quarkonium has been integrated out \cite{Blaizot:2018oev}. In the limit of small momentum transfer $q$, it reduces to a dipolar interaction, as we shall discuss shortly.  In Eq.~(\ref{eq:width}), $|s\rangle$ is the wave function of the singlet and $E_{\rm s}$ its binding energy. The ket  $|{\rm o},{\bf p}\rangle$ is the state of a heavy quark pair in an octet state with energy $E_{{\bf p}}^{{\rm o}}=\frac{p^2}{M}$. In the large $N_c$ limit the octet potential is zero, therefore, the wave function of an octet can be approximated by a plane wave. As a consequence, the expectation value $\langle{\rm s}|\sin\left(\frac{{\bf q}\hat{{\bf r}}}{2}\right)|{\rm o},{\bf p}\rangle$ is easily obtained as a Fourier transform of the wave function of the singlet bound state. The last ingredient needed to compute the decay width is the correlator $\Delta^{>}(w,{\bf q})$ (see Eq.~(\ref{correlators})). We shall estimate it using the Hard Thermal Loop (HTL) approximation \cite{Pisarski:1988vd,Braaten:1989mz,Frenkel:1989br}.
\begin{equation}
\Delta^{>}(w,{\bf q})=\frac{{\rm e}^{w/T}}{{\rm e}^{w/T}-1}\sigma(w,{\bf q})\,,
\end{equation}
where
\begin{equation}
\sigma(\omega,{\bf q})=\frac{2{\rm Im}\Pi_{L}(\omega,{\bf q})}{(q^{2}+{\rm Re}\Pi_{L}(\omega,{\bf q}))^{2}+({\rm Im}\Pi_{L}(\omega,{\bf q}))^{2}}\,,
\end{equation}
and
\begin{equation}
\Pi_{L}(\omega,{\bf q})=m_{D}^{2}(T)\left(1-\frac{\omega}{2q}\ln\left(\frac{\omega+q+i\epsilon}{\omega-q+i\epsilon}\right)\right)
\end{equation}
is the longitudinal component of the gluon polarization tensor. It is proportional to the 
  Debye mass $m_D$, given in leading order perturbation theory by 
 \begin{equation}
m_{D}(T)=\sqrt{\frac{4\pi\alpha_{s}(N_{c}+T_{F}N_{F})}{3}}T\,,
\label{eq:md}
\end{equation}
where $N_c=3$, $T_F=\frac{1}{2}$ and $N_F$ is the number of flavours that can be considered light when computing $m_D$ (we take $N_F=3$). A plot of this Debye mass as a function of the temperature, as obtained with different scales for the running coupling constant $\alpha_s$, is given  in Fig.~\ref{fig:md}. One can see that, for this range of temperatures, $m_D\simeq 2.5 T$, a result that is not truly in the perturbative regime. Thus we should regard our use of the HTL result as a convenient phenomenological guide, providing us with a reasonable approximation for  the overall momentum and frequency dependence of the longitudinal tensor, while  $m_D$, to which it is proportional, could be viewed as a phenomenological parameter. 

\begin{figure}
\includegraphics[scale=0.65]{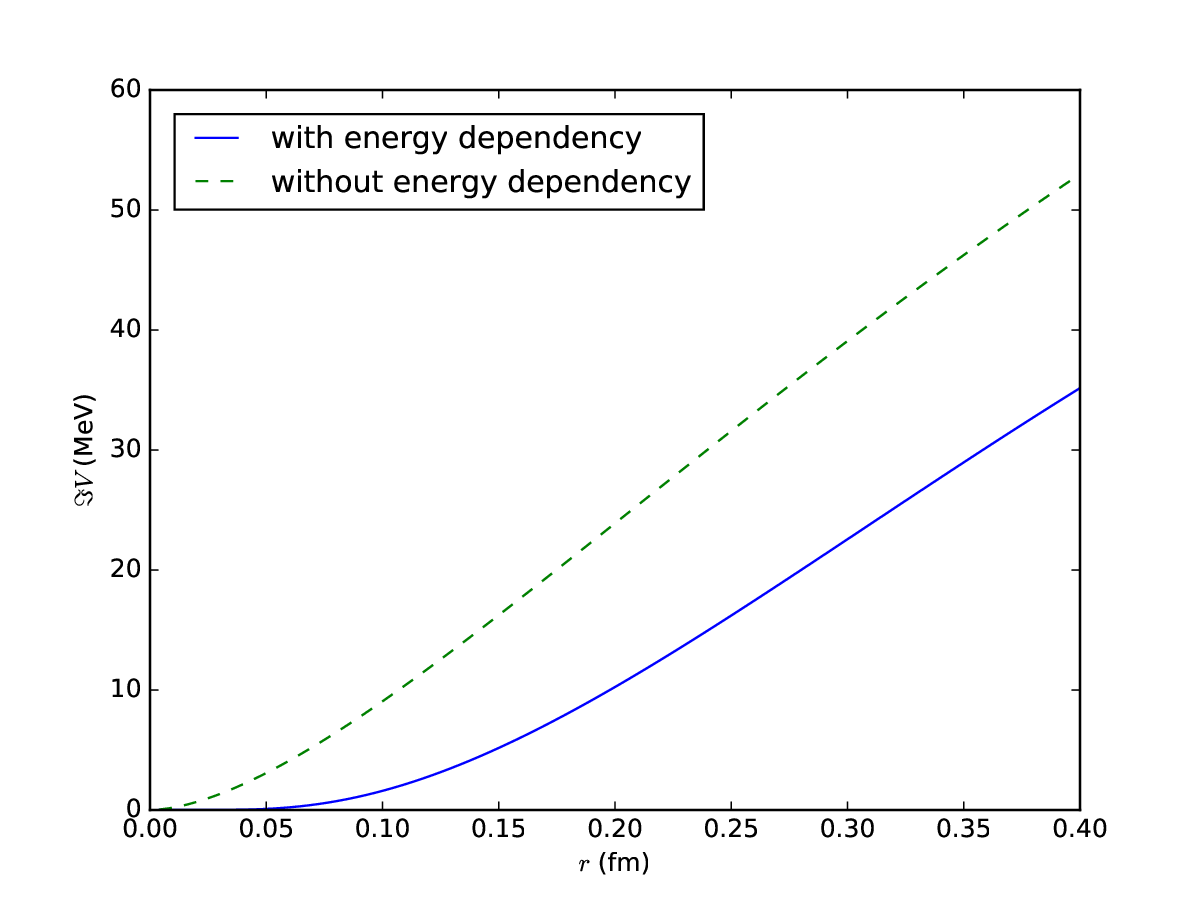}
\caption{Comparison between the imaginary part of the potential at $T=250\,\text{MeV}$ obtained from Eq. (\ref{eq:width}) (with energy dependency) or from Eq.~(\ref{eq:widthgl})  (without energy dependency). Note that the effect of energy dependence is sizeable and amounts to typically a factor 2 reduction. Picture taken from \cite{Blaizot:2018oev}.}
\label{fig:ImV}
\end{figure}

If one ignores the  gap between singlets and octets,
which may be legitimate when the temperature is high enough, one obtains the simpler expression 
\begin{equation}
\Gamma_{s}^0=8\pi\alpha_sC_{F}\int_{{\bf q}}\Delta^{>}(0,{\bf q})\langle{\rm s}|\sin^2\left(\frac{{\bf q}\hat{{\bf r}}}{2}\right)|{\rm s}\rangle\,,
\label{eq:widthgl}
\end{equation}
which we refer to as the  static limit \footnote{Here we refer to the static limit as the limit $T\gg\Delta E$. Note that in other context the static limit is understood as the limit in which the heavy quark mass is infinite.}. This expression  (\ref{eq:widthgl}) can be read as twice the imaginary part of the potential, as originally computed in \cite{Laine:2006ns}. One may then interpret Eq.~(\ref{eq:width}) in similar terms, as an energy dependent imaginary potential. This energy dependence, present in the exponential factor as well as in the correlator,  plays a crucial role at low temperature since it ensures that the fluctuation-dissipation theorem is fulfilled. The impact of this energy dependence of the imaginary potential was studied numerically in \cite{Blaizot:2018oev}, and is illustrated in Fig. \ref{fig:ImV}. One can see that ignoring this energy dependence typically increases the decay width by a factor 2. The purpose of this note is to quantify this effect in the more realistic situation of an expanding plasma. We note that the importance of the energy gap was recognized in early studies of quarkonium interactions with matter \cite{Kharzeev:1994pz,Kharzeev:1995ij}. 

Before closing this subsection, we note that the energy dependence  in Eq. (\ref{eq:width}) originates from the energy gap between singlets and octets. We determine the singlet binding energy  by solving the Schr\"{o}dinger equation
\begin{equation}
H_Q|{\rm s}\rangle=E_{\rm s}|{\rm s}\rangle\,, \qquad 
H_Q=\frac{p^2}{M}+V_{\rm s}(r)\,.
\label{eq:HQ}
\end{equation}
The potential $V_{\rm s}$ is a screened potential which will be specified shortly. At this point we just comment on the interplay between screening and decay rate:
\begin{itemize}
\item For simplicity, we set the origin of energies as that of the lowest energy octet state. Then the absolute value of the gap between a singlet state and the octet is equal to the binding energy of the singlet.  Then, for the model to be valid, the condition $|E_{\rm s}|\gg\Gamma_{\rm s}$ must be fulfilled. This is equivalent to saying that the time scale for which thermal effects become important must be much larger than the inverse of typical energy difference between states. If this is not so, we cannot consider thermal effects as a perturbation using multiple-scale analysis (see discussion before Eq.~(\ref{eq:multi})), which is crucial to get to the simple form of Eq.~(\ref{eq:modfin}).
\item The decay width increases rapidly as the gap decreases. Therefore, the phenomena of screening and decay due to the collisions influence each other. As the temperature increases, so does screening. Screening reduces the energy gap between singlets and octets, and this results in an increase of the decay width. 
\end{itemize}

Because of this strong interplay, it is important to have a consistent and simultaneous treatment of both phenomena.

\subsection{Relation with the dipole approach}

At this point we wish to make a comment on the dipole approximation used in most EFT calculations based on the assumption of a clean separation of scales. 
In the (strict)  limit $\frac{1}{r}\gg m_D \gg \Delta E$,  Eq.~(\ref{eq:widthgl}) reduces to 
\begin{equation}
\Gamma_{\rm s}^0\sim 2\pi\alpha_s(\mu_T)C_{F}\int_{{\bf q}}q^iq^j\Delta^{>}(0,{\bf q})\langle s|r^ir^j|s\rangle\,.
\label{eq:Gammasimp}
\end{equation}
This expression is plagued by an ultraviolet divergence whose origin can be found in the HTL approximation used for the gluon propagator. This approximation is only valid for $q\lesssim m_D$, which is in principle compatible with our assumption. However, in expanding the sine function, assuming that $q r\ll 1$, we have  produced a factor $q^i q^j$ whose magnitude is not controlled anymore by any factor. The HTL correlator results then integrated over momenta where it is not valid. This is the origin of the divergence. This divergence can be coped with by introducing a cutoff $\Lambda$ in the $q$-integration, using the HTL approximation of the correlator for $q<\Lambda$ and the full one-loop correlator for $q>\Lambda$. Since the divergence is logarithmic, the dependence on the cutoff  eventually disappears, leaving contributions proportional to $\ln(T/m_D)$. We should note however that such a logarithm, while positive in the strict perturbative regime where $T\gg m_D$, may turn negative in the temperature range probed by present experiments (where the coupling constant  may remain of order unity), leading to unphysical values for the decay rate. 

An alternative to the procedure outlined above is to use a better approximation for the correlator  $\Delta^{>}(0,{\bf q})$ in 
Eq. (\ref{eq:Gammasimp}). We can rewrite this equation  as \cite{Brambilla:2016wgg,Brambilla:2017zei}
\begin{equation}\label{eq:diffusioncste}
\Gamma_{\rm s}^0\sim\frac{\pi\alpha_s(\mu_T)}{3N_c}\int\,dt\langle\nabla^iA_0^A(t,\mathbf{0})\nabla^iA_0^A(0,\mathbf{0})\rangle \langle s|r^2|s\rangle\sim \langle s|r^2|s\rangle\kappa,
\end{equation}
with $\kappa$ a momentum diffusion coefficient  \cite{Moore:2004tg}. This formula relates this diffusion coefficient to the correlator of electric field fluctuations, and this can be calculated beyond the HTL approximation, including possibly non perturbative effects \cite{CasalderreySolana:2006rq}. A  detailed discussion of these issues  can be found in \cite{Akamatsu:2020ypb}. Here we note that in the important regime where the various scales are overlapping, a static diffusion constant is in any case not enough, since it ignores the frequency dependence which we claim is important. 

We now present details of the  implementations of the model and examine two scenarios to fix the parameters. For lack of a better terminology, we shall refer to these scenarios as the  perturbative scenario and the lattice scenario. It should be understood however that the perturbative scenario departs from strict perturbation theory (as we have already alluded to) and the lattice scenario is not  a lattice calculation.

\subsection{The perturbative scenario}
We obtain the binding energy and the wave-functions needed to compute $\Gamma_{\rm s}$ and $\Gamma_{\rm s}^0$ by solving numerically the Schr\"{o}dinger equation,  using the algorithm described in \cite{Lucha:1998xc}. We focus on $\Upsilon(1S)$ because it is only for this state that, within this scenario, the condition that the binding energy is much bigger than the decay width is fulfilled. This condition is needed for Eq.~(\ref{eq:modfin}) to be valid. In the case of $\Upsilon(2S)$, the decay width is of the same magnitude as the binding energy at the lowest temperature that we consider. We need to specify first the hamiltonian $H_Q$ in Eq. (\ref{eq:HQ}).  For $M$, we take as definition a naive version of the 1S of mass \cite{Hoang:1999ye}, which consists in setting $M=\frac{M_{\Upsilon(1S)}}{2}\sim 4730\,\textit{MeV}$. We think that, given the accuracy we are aiming at, this is a reasonable option.

For the real part of the potential,  we use a screened Yukawa potential
\begin{equation}\label{screenV}
V_s(r)=-\frac{C_F\alpha_s(\mu_r)e^{-m_D(T,\,\mu_T)r}}{r}\,,
\end{equation}
where $\mu_r$ and $\mu_T$ are (independent) subtraction points. The scale $\mu_r$ is related to the exchange of gluons between the heavy quark and the antiquark and enters the running coupling $\alpha_s$ in Eqs. (\ref{eq:width}) and (\ref{screenV}). A natural choice would be $\mu_r\sim\frac{1}{r}$, but this would lead to difficulties at large distance $r$ in the numerical   implementation of the running coupling.  Therefore we fix $\mu_r\sim\frac{1}{a_0}$, where $a_0$ is the Bohr radius, which we define with the following self-consistent equation
\begin{equation}
\frac{1}{a_{0}}=\frac{MC_{F}\alpha_{s}(1/a_{0})}{2}\,,\qquad a_0=0.149\,\textit{fm}\,.
\end{equation}
 The scale $\mu_T$ is the scale at which $\alpha_s$ is evaluated when computing the Debye mass using Eq. (\ref{eq:md}). The Debye mass encodes the influence of the particles with a typical energy of order $\pi T$, which is therefore a natural choice for the value of $\mu_T$. Finally, we use the leading order $\beta$-function for the running of the coupling constant, with $\Lambda_{\rm QCD}=250\,\text{MeV }$ and  $N_{F}=3$.

\begin{figure}
\includegraphics[scale=0.75]{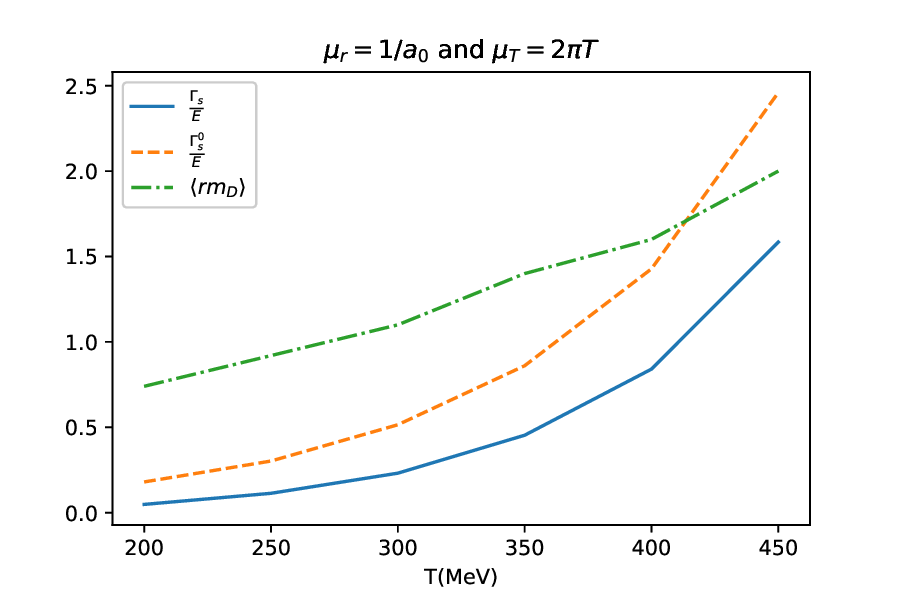}
\caption{Comparison of the dimensionless quantities $\langle rm_D\rangle$, $\frac{\Gamma_{\rm s}}{E}$ and $\frac{\Gamma^0_{\rm s}}{E}$. The subtraction points are $\mu_r=\frac{1}{a_0}$ and $\mu_T=2\pi T$.}
\label{fig:P1_r2}
\end{figure}

Fig.~\ref{fig:P1_r2} illustrates various criteria that are often considered for the melting of the quarkonia. The one based on screening states that the bound state disappears when $<r>$, the average size of the bound state, becomes comparable to the Debye screening length $\sim 1/m_D$. Thus one expects screening alone to affect the survival of bound states when $<r m_D >$ becomes of order unity. The other criterion focuses on collisions and considers that the bound state disappears when the decay width $\Gamma_{\rm s}$  becomes of comparable magnitude as the binding energy $E_{\rm s}$, that is when $\Gamma_{\rm s}/E_{\rm s}$ becomes of order unity.  We note that both $<rm_D>$ and $\Gamma_{\rm s}/E_{\rm s}$ are increasing functions of the temperature. Fig.~\ref{fig:P1_r2} also  illustrates the influence of the gap by  comparing the ratios ${\Gamma_{\rm s}}/{E}$ and  ${\Gamma^0_{\rm s}}/{E_{\rm s}}$. We remind that the model is valid as long as this ratio is well below $1$.

\begin{figure}
\includegraphics[scale=0.5]{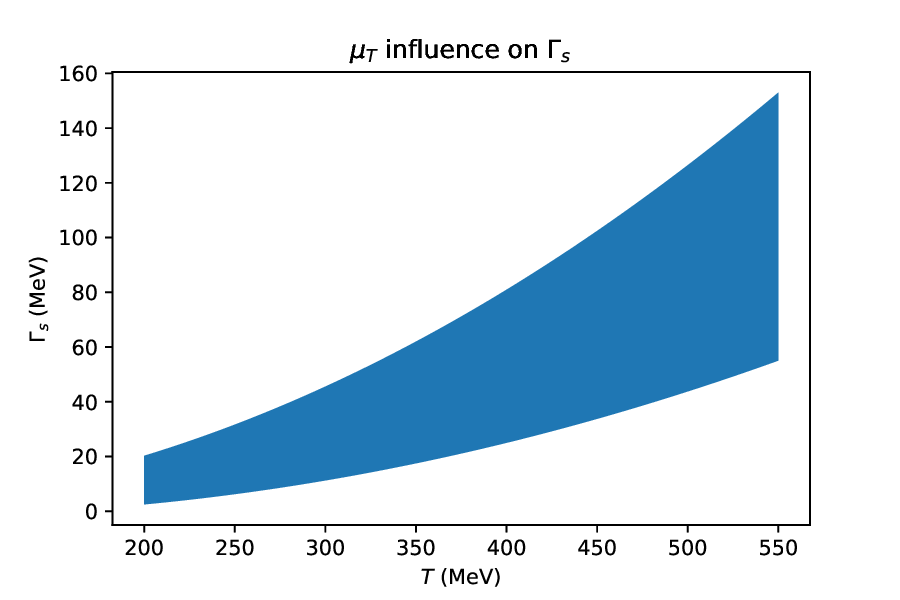}\includegraphics[scale=0.5]{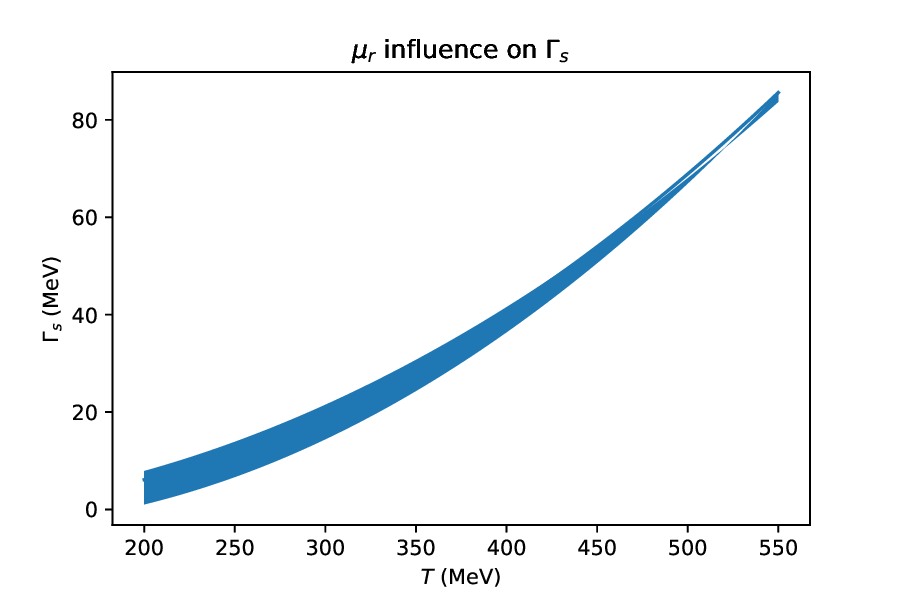}
\caption{These plots illustrate the influence of $\mu_T$ (left) and $\mu_r$ (right) on $\Gamma_s$. In the left plot,   $\mu_r=\frac{1}{a_0}$ while $\mu_T=\pi T, 2\pi T, 4\pi T$. In the right plot $\mu_T=2\pi T$, and  $\mu_r=\frac{1}{2a_0}, \frac{1}{a_0}, \frac{2}{a_0}$.}
\label{fig:inmut}
\end{figure}

\begin{figure}
\includegraphics[scale=0.5]{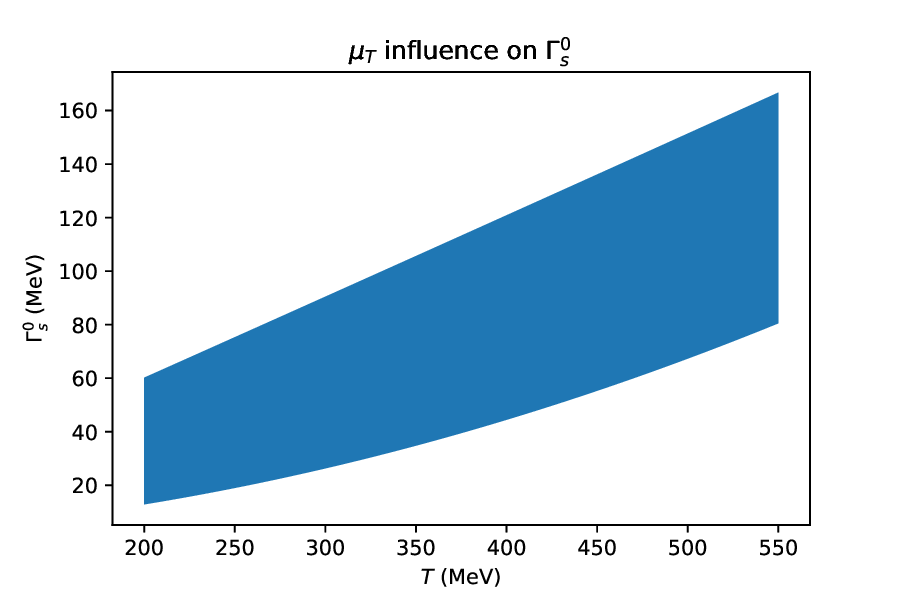}\includegraphics[scale=0.5]{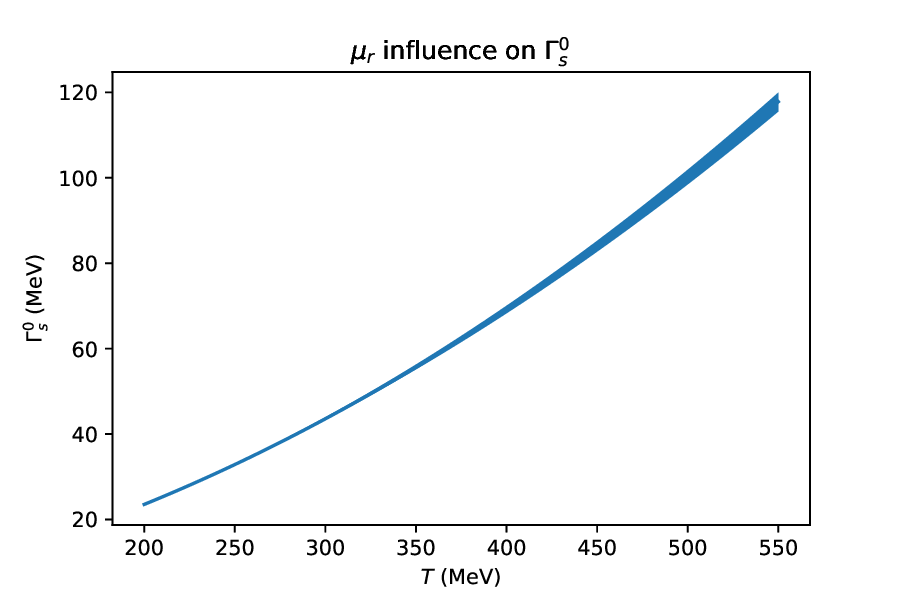}
\caption{Same as Fig. \ref{fig:inmut} but for $\Gamma_{\rm s}^0$.}
\label{fig:inmutgl}
\end{figure}

Consider now the uncertainty that results from the choices of $\mu_r$ and $\mu_T$ in the computation of the decay width. We only computed $\Gamma_{\rm s}$ and $\Gamma^0_{\rm s}$ for a finite set of temperatures  (increasing $T$ from 200 to 450 MeV by steps of 50 MeV) to save on computational cost. However, we have observed that the following function provides a good fit of our results,
\begin{equation}
\Gamma_{\rm s}\sim aT+bT^2\,.
\label{eq:gammafit}
\end{equation}
There is no theoretical reason for choosing this fit function apart from the fact that at very high energies one expects the decay to grow linearly with the temperature (neglecting the running of the coupling). This fit function will be used later when we calculate the $R_{AA}$ for the expanding system. Here this is used as an interpolation formula to estimate the error associated with the choices of the various subtraction scales. Figs.~\ref{fig:inmut} and \ref{fig:inmutgl} illustrate the sensitivity of respectively $\Gamma_{\rm s}$ and $\Gamma^0_{\rm s}$ to variations of $\mu_r$ and $\mu_T$ around nominal values, chosen to be respectively $1/a_0$ and $2\pi T$.  Clearly the uncertainty associated with variations of $\mu_T$ is bigger than that associated to  changes in $\mu_r$. Overall, we observe that $\Gamma^0_{\rm s}>\Gamma_{\rm s}$, as expected.
\subsection{The lattice scenario}
\begin{figure}
\includegraphics[scale=0.5]{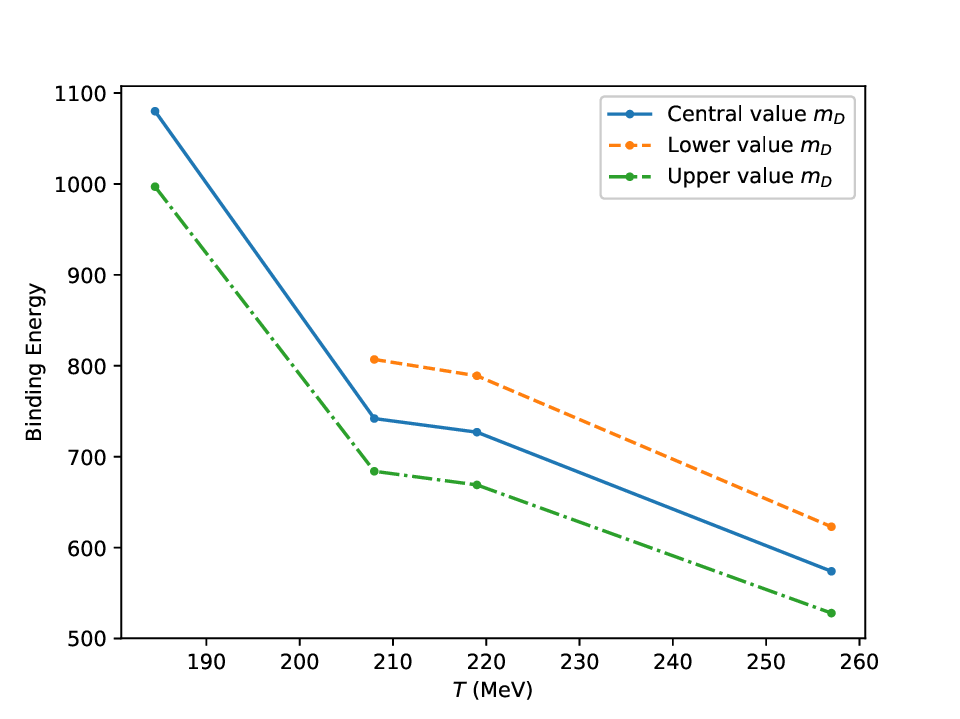}\includegraphics[scale=0.5]{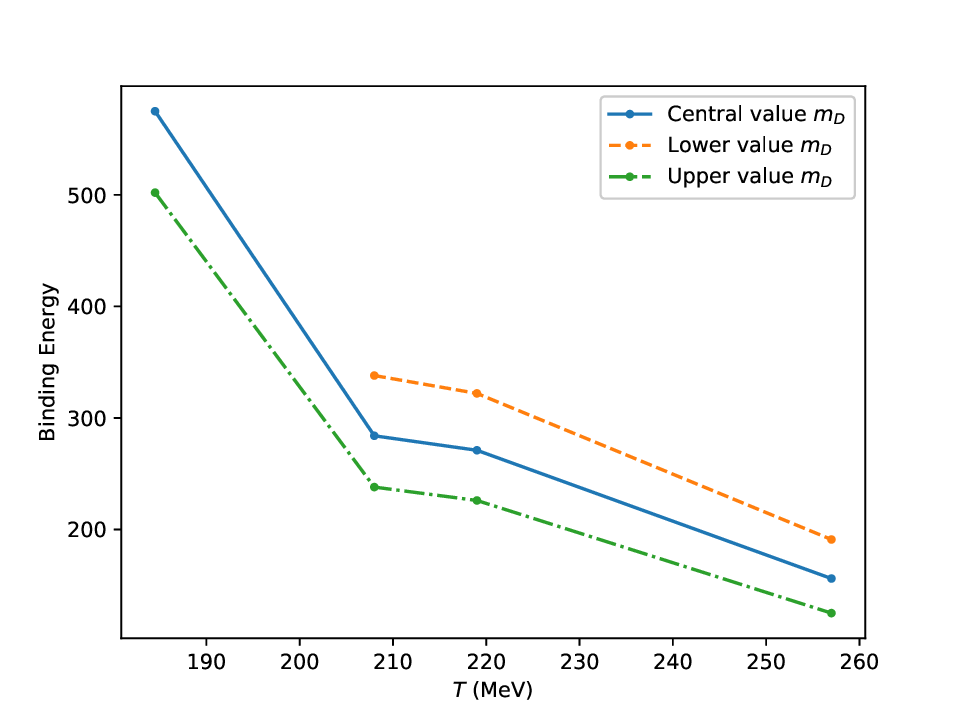}
\caption{Binding energy of $\Upsilon(1S)$ (left) and  $\Upsilon(2S)$ (right) in the strongly coupled scenario}
\label{fig:st1be1s}
\end{figure}

\begin{table}
\begin{tabular}{|c|c|}
\hline
$T$ (MeV) & $m_D$ (MeV) \\
\hline
$184.5$ & $317\pm 25$\\
$208$ & $437.5\pm 29.5$\\
$219$ & $445\pm 30$\\
$257$ & $534\pm 33$\\
\hline
\end{tabular}
\caption{Debye mass as a function of temperature as extracted from the fit in \cite{Lafferty:2019jpr}.}
\label{tab:md}
\end{table}
As an alternative possibility to fix the basic ingredients of the model, we shall rely on  the lattice data from a recent computation \citep{Lafferty:2019jpr}. We  refer to this second strategy as the ``lattice scenario''.  As in the perturbative scenario, the binding energy and the wave function of the quarkonia are obtained by solving a Schr\"{o}dinger equation. However, now we take as heavy quark mass the same as in \cite{Lafferty:2019jpr}, $M=4882\,\text{MeV}$, slightly larger than that of the perturbative scenario. As for the real part of the potential, it is given by 
\begin{equation}\label{latticeVs}
V_{s}(r)=-\frac{\alpha e^{-m_{D}r}}{r}-\sigma re^{-m_{D}r}\left(1+\frac{2}{m_{D}r}\right)\,,
\end{equation}
where $\alpha$, $\sigma$ and $m_D$ are obtained from the fit performed in \cite{Lafferty:2019jpr}. The values of $m_D$  differ from those in the perturbative scenario, and they are given at a different set of  temperatures. These values are listed in table \ref{tab:md} together with the corresponding temperatures at which they are determined. The definition of the potential is such  that  $V_{\rm s}(\infty)=0$, i.e. states become unbound at $E=0$. We identify therefore  $-E_{1S}$ as the energy gap between the 1S state and the octets, with $E_{1S}$ the binding energy. It may happen that, at low temperature, the solution of the Schr\"{o}dinger equation with the potential (\ref{latticeVs}) yields a binding energy larger  than that obtained at $T=0 $ and computed as $2M_B-M_\Upsilon$, where $M_B$ is the  mass of a $B$ meson and $M_\Upsilon$ the mass of the $\Upsilon(1S)$. We discard such cases as unphysical. This is somewhat similar to what is done in Eq.~(3.5) of  \cite{Islam_2021}. This is why  no result is given in Fig.~\ref{fig:st1be1s} for $T=184.5$ MeV and the lowest value of $m_D$.

\begin{figure}
\includegraphics[scale=0.5]{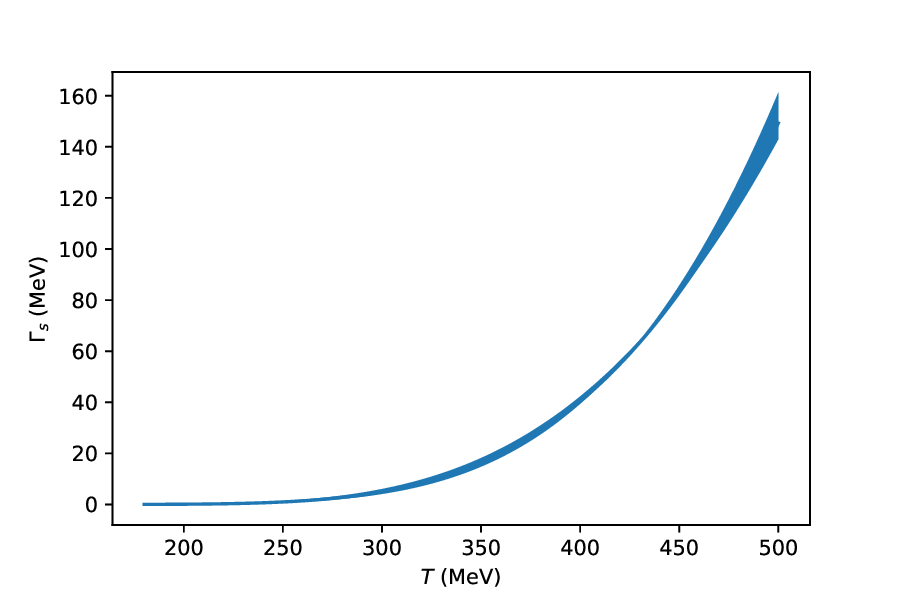}\includegraphics[scale=0.5]{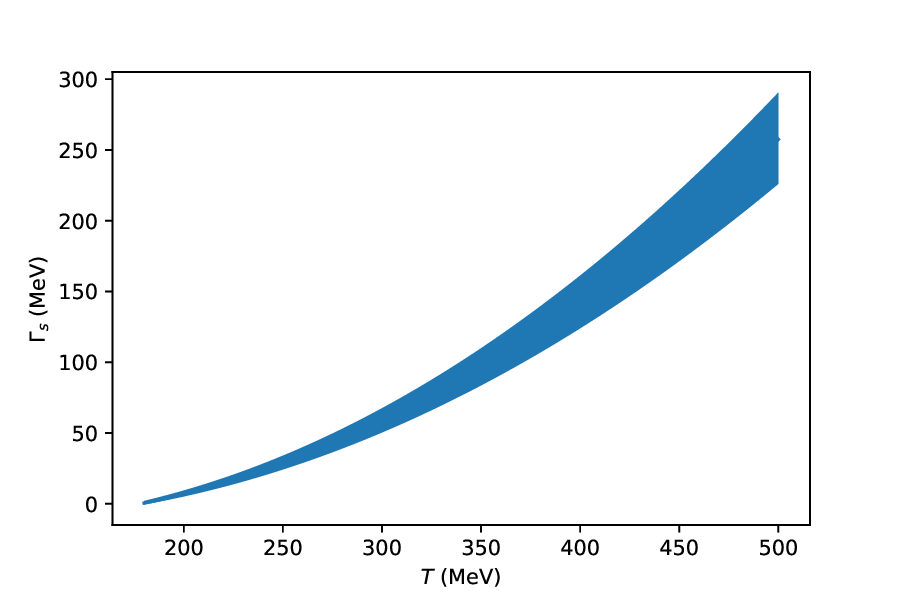}
\caption{The decay width $\Gamma_{\rm s}$ of $\Upsilon(1S)$ (left) and $\Upsilon(2S)$ (right)  in the lattice scenario.}
\label{fig:st11s}
\end{figure}

\begin{figure}
\includegraphics[scale=0.5]{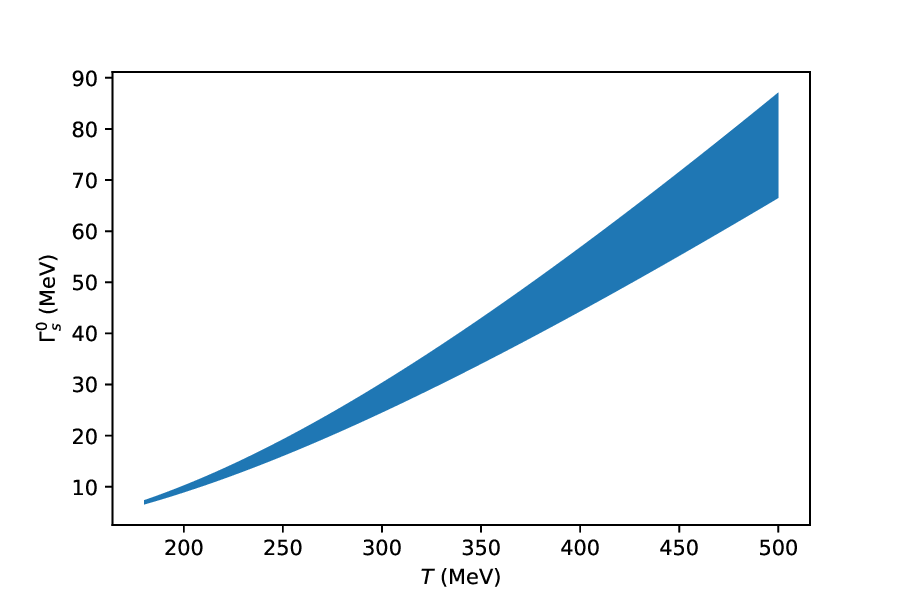}\includegraphics[scale=0.5]{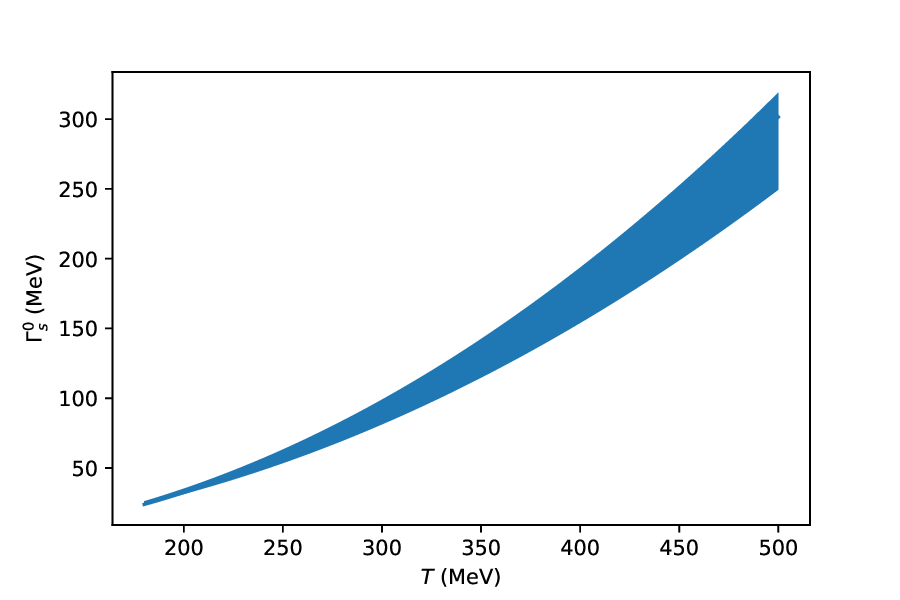}
\caption{The decay width $\Gamma_{\rm s}^0 $ of  $\Upsilon(1S)$ (left) and $\Upsilon(2S)$ (right) in the lattice scenario.}
\label{fig:st11sgl}
\end{figure}

The decay width is obtained from  Eq. (\ref{eq:width}), with now the values of $m_D$ listed in Table~\ref{tab:md},  and $\alpha_s={\alpha}/{C_F}$ with  $\alpha$  taken from \cite{Lafferty:2019jpr}. With these parameters, it becomes possible to study both  $\Upsilon(1S)$ and $\Upsilon(2S)$ since for both states the condition $E_{\rm s}> \Gamma_{\rm s}$ is  fulfilled. The results of the calculation are displayed in Fig.~\ref{fig:st1be1s} for the binding energy and in Figs.~\ref{fig:st11s} and \ref{fig:st11sgl} for the decay widths $\Gamma_{\rm s}$ and $\Gamma^0_{\rm s}$ respectively. 

In the case of the $\Upsilon(1S)$, the decay width is well fitted by the formula
\begin{equation}
\Gamma_{\rm s}\sim \tilde{a}Te^{-\frac{\tilde{b}}{T}}
\label{eq:gammafit2}
\end{equation}
motivated in part by the fact that  at high temperatures the decay width is linear with the temperature while at small temperatures we expect an exponential  suppression due to the gap. For the case of $\Upsilon(2S)$, however, we found that Eq. (\ref{eq:gammafit}) works better as a fit function. These formulae are used in the analysis presented in  Figs.~\ref{fig:st11s} and \ref{fig:st11sgl}. We note that the fact that we use the same fit function for $\Upsilon(2S)$ as we did for $\Upsilon(1S)$ in the perturbative scenario does not imply that our computation for $\Upsilon(2S)$ is perturbative. 

\begin{table}
\caption{Parameters obtained by fitting the decay width in the three cases (from S1 to S3) described in the text. The parameters $\tilde{a}$ and $\tilde{b}$ correspond to the fit of the function of Eq.~(\ref{eq:gammafit2}) for the case of $\Upsilon(1S)$. The parameters $a$ and $b$ correspond to the fit of the function of Eq.~(\ref{eq:gammafit}) for the case of $\Upsilon(2S)$.}
\label{tb:st}
\begin{tabular}{|c|c|c|c|c|}
\hline
Label & $\tilde{a}$ & $\tilde{b}\,(\textit{MeV})$ & $a$ & $b\,(\textit{MeV})^{-1}$ \\
\hline
S1 & $22.9$ & $217\cdot 10$ & $-0.285$ & $0.00160$ \\
S2 & $36.9$ & $237\cdot 10$ & $-0.258$ & $0.00142$ \\
S3 & $15.3$ & $199\cdot 10$ & $-0.309$ & $0.00178$ \\
\hline
\end{tabular}
\end{table}

The biggest source of error in this lattice scenario comes from the value of $m_D$. We consider three cases, referred to as S1, S2 and S3. The case S1 uses the central value of $m_D$. The case S2 and S3 correspond respectively to the  the lowest (S2) or the largest (S3) values of $m_D$ that are compatible with the error given in \cite{Lafferty:2019jpr}.  The resulting fit  parameters are  given in table \ref{tb:st}. Using these parameters we obtain the decay widths of $\Upsilon(1S)$ and $\Upsilon(2S)$ shown in Fig. \ref{fig:st11s}. Note that these figures contain an extrapolation to temperatures that are larger than those available in the lattice calculations (which are limited to $T\le 260$ MeV). 

The results we have obtained for the binding energy and the decay width can be compared to  recent results in the literature. The binding energies that we have obtained for $\Upsilon(1S)$ and $\Upsilon(2S)$ are shown in Figs. \ref{fig:st1be1s}. For $\Upsilon(1S)$ these are qualitatively similar and of the same order of magnitude as those in \cite{Thakur:2020ifi,Srivastava:2018vxp,Thakur:2013nia,Mocsy:2007jz,Hasan:2020iwa}. For $\Upsilon(2S)$, our results are qualitatively similar to those in \cite{Thakur:2020ifi,Srivastava:2018vxp} but about  five times smaller than those  found in \cite{Mocsy:2007jz}, in which the decay width is not based in the HTL model but rather on the computation in \cite{Kharzeev:1995ju}. Concerning the decay width, we note that, in general, we obtain results for the  $\Upsilon(1S)$ that are of the same order of magnitude as those in \cite{Thakur:2020ifi,Srivastava:2018vxp,Thakur:2013nia,Mocsy:2007jz,Hasan:2020iwa}, however, the qualitative behavior is different since our results are more suppressed at small temperatures. The reason is that we take into account the suppression due to the energy gap between singlet and octet states, which was not taken into account in the other studies. We observe a similar behaviour in the case of $\Upsilon(2S)$, with the difference that the effect of the gap is much less pronounced here. 

We have repeated the analysis for $\Gamma^0_{\rm s}$ (Fig.~\ref{fig:st11sgl}). The values that we obtained for $\Gamma^0_{\rm s}$ are substantially larger than those of $\Gamma_{\rm s}$, as it should, as long as we remain in the temperature region used in \cite{Lafferty:2019jpr}. We observe, however,  that at large temperatures $\Gamma_{\rm s} $ in Fig. \ref{fig:st11s} is actually larger than $\Gamma^0_{\rm s}$ in Fig. \ref{fig:st11sgl}. We view this as an artefact of extrapolating the results using Eq. (\ref{eq:gammafit2}) far beyond the region in which we performed the fit. This indicates that we have to be cautious when interpreting our results in the lattice   scenario for collisions that probe high temperatures.

\section{Phenomenological implications}
\label{sec:pheno}

We turn now to the main objective of this paper, which is to estimate the quantitative impact of the expansion on the previous results. To do so, we shall provide a crude estimate of  the observable  $R_{AA}$. We assume that the  quarkonium state, initially in a state $n$ starts interacting with the plasma at some (small) finite time which we choose to be $t_0=0.6\,\textit{fm}$ (value taken from \cite{Alberico:2013bza}). The survival probability is then obtained from the rate equation (see Eq. (\ref{eq:modfin}))
\begin{equation}
\frac{\rmd p_n}{\rmd t}=-\Gamma(T(t))p_n(t)\,,
\label{eq:Raa}
\end{equation}
with the initial condition $p_n(t_0)=1$.  The survival probability of a given quarkonium state is $S=p_n(t_f)$, where $t_f$ is the time spent by the quarkonium in the plasma. Its value depends on the scenario considered and it will be specified shortly. The above calculation assumes that the temperature evolves slowly with time so that the interaction with the plasma can be treated in an adiabatic approximation. 
That is, one assumes that the quarkonium interacts with the medium as in the static case, with the medium characterized by its instantaneous temperature $T(t)$. 

For simplicity, we assume that the center of mass of the quarkonium does not move in the transverse plane. Thus the survival probability depends only on the time spent in the plasma, and on the local conditions in which it evolves \cite{Blaizot:1988hh}. These depend  on the impact parameter, $\b$, and on the point in the transverse plane, $\s$, (with respect to the collision axis) in which the quarkonium is produced. We set the origin of coordinates in the transverse plane to coincide with the center of one of the nucleus. The initial temperature of the medium, denoted by $T_0(\b,\s)$,  is related to the local energy density $\varepsilon$ by the standard relation $\varepsilon\sim T^4$. The energy density itself is taken to be proportional to the surface density of participants, which we estimate as a function of $\b$ and $\s$ using a Glauber model \cite{Blaizot:1988ec}. Thus we write
\begin{widetext}
\begin{equation}
T_0(\b,\s)=T_0(0,0)\left(\frac{T_A(\s)\left[1-\left(1-\frac{\sigma T_A(\mathbf{s}-\mathbf{b})}{A}\right)^A\right]}
{T_A(0)\left[1-\left(1-\frac{\sigma T_A(0)}{A}\right)^A\right]}\right)^{1/4}\,,
\end{equation}
\end{widetext}
where $T_A$ is the overlap function in the Glauber model, $A=207$ and (in the type of collisions considered here) $\sigma=70\,\text{mb}$ \cite{Loizides:2014vua}.
We take $T_0(0,0)=475\times 1.05\,\text{TeV}$ for $\sqrt{s}=5.02\,\text{TeV}$ collisions at LHC. This number is obtained by using the standard value of \cite{Alberico:2013bza} and taking into account that the temperature increases by about $5\%$ when going from $2.76\,\text{TeV}$ collisions to $5.02\,\text{TeV}$ \cite{Alqahtani:2020paa}. 
Regarding the time dependence of the temperature, we use  Bjorken hydrodynamics, and ignore the transverse expansion,
\begin{equation}
T(t,\b,\s)=T_0(\b,\s)\left(\frac{t_0}{t}\right)^\frac{1}{3}\,.
\end{equation}

Now we have all the ingredients to compute $R_{AA}$. According to the optical Glauber model, $R_{AA}$ for a given impact parameter is computed as
\begin{equation}
R_{AA}(\b)=\frac{\int\,d^2\s T_A(\s)T_A(\mathbf{s}-\mathbf{b})S(\b,\s)}{\int\,d^2\s T_A(\s)T_A(\mathbf{s}-\mathbf{b})}\,,
\end{equation} 
where $S(\b,\s)$ is the survival probability. 

\subsection{Perturbative scenario}
\begin{figure}
\includegraphics[scale=0.5]{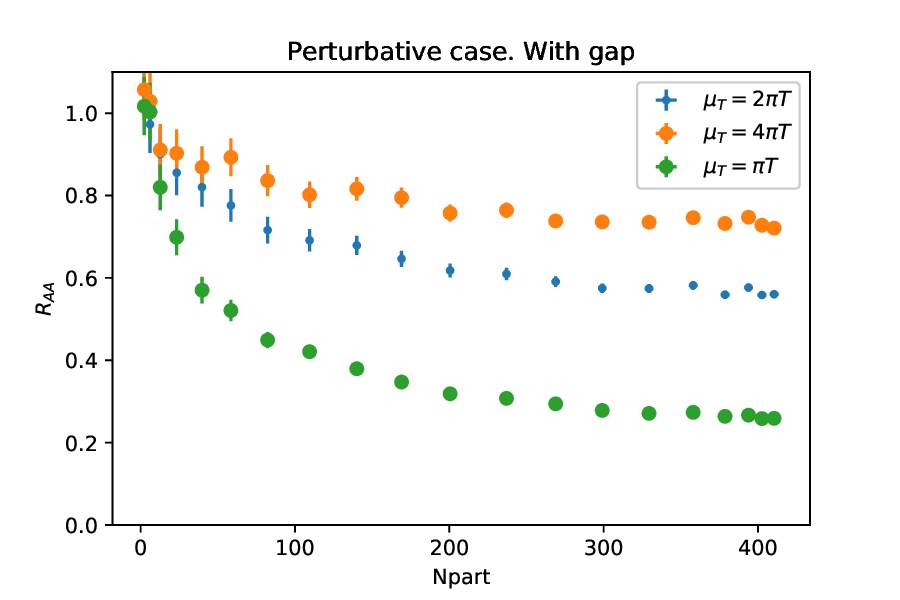}\includegraphics[scale=0.5]{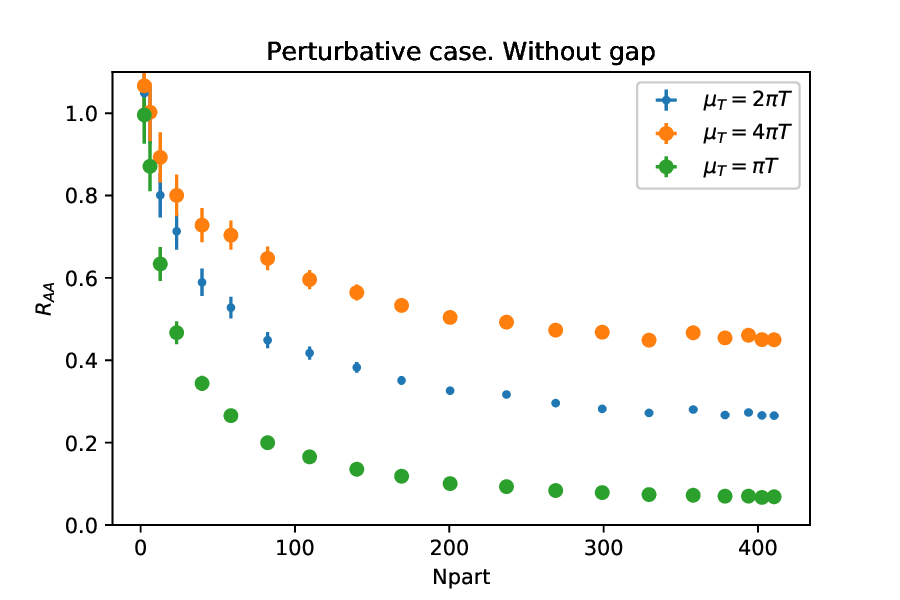}
\caption{$R_{AA}$ of $\Upsilon(1S)$ obtained with the perturbative scenario against the number of participants, with (left) and without (right) the energy dependence of the decay rate. The error quoted is due to Monte Carlo integration.}
\label{fig:htlraavsnpart}
\end{figure}
The survival probability can be computed analytically if we use for the decay width the approximate expression (\ref{eq:gammafit}). One gets then
\begin{equation}
S_{(1S)}(\b,\s)={\rm e}^{-1.5aT_{0}(\b,\s)t_{0}\left(\left(\frac{T_{0}(\b,\s)}{T_{f}}\right)^{2}-1\right)-3bT_{0}(\b,\s)^{2}t_{0}\left(\frac{T_{0}(\b,\s)}{T_{f}}-1\right)}\,.
\label{eq:suv1}
\end{equation}
We can use this formula to compute $R_{AA}$. In this case, we set $T_f=200\,\text{MeV}$, arguing that  a perturbative computation is only valid for temperatures above that of the phase transition. We are aware that physics at lower temperatures might modify the survival probability. However, we ignore those effects. At the moment, our aim is not to obtain a state-of-the-art phenomenological prediction, but to highlight the importance of the energy gap between singlets and octets. 

 In Fig.~\ref{fig:htlraavsnpart} we plot the results obtained by applying this formula, taking the dependence on $\mu_T$ as a measure of our theoretical uncertainty. Although the uncertainty due to $\mu_T$ is quite significant, we can clearly see the difference between $\Gamma_{\rm s}$ and $\Gamma^0_{\rm s}$. We see that taking into account the gap increases $R_{AA}$ by typically the same factor $\sim 2$ as in the non expanding case. We note that in Fig.~\ref{fig:htlraavsnpart}, and in the rest of the figures, we have plotted our results in the full range of centralities although our model is only valid when $E\gg \Gamma_{\rm s}$.

\subsection{The lattice scenario}
The computation is completely analogous to what was already explained in the perturbative scenario. The only difference is that,  in the case of $\Upsilon(1S)$ we used a fitting function of the type of Eq.~(\ref{eq:gammafit2}) instead of Eq.~(\ref{eq:gammafit}). In the case of a decay width which follows Eq.~(\ref{eq:gammafit2}) we can also obtain an analytic solution for $S$
\begin{equation}
S(\b,\s)=e^{-\frac{3\tilde{a}T_{0}(\b,\s)^{3}t_{0}}{\tilde{b}^{2}}\left(e^{-\frac{\tilde{b}}{T_{0}(b,s)}}\left(1+\frac{\tilde{b}}{T_{0}(\b,\s)}\right)-e^{-\frac{\tilde{b}}{T_{f}}}\left(1+\frac{\tilde{b}}{T_{f}}\right)\right)}\,.
\end{equation} 
In the case of $\Upsilon(2S)$, we can also apply Eq.~(\ref{eq:suv1}). Another difference with respect to the perturbative scenario is that now we use $T_f=180\,\text{MeV}$. As we explained before, for temperatures smaller than this value, the fit obtained in \cite{Lafferty:2019jpr} gives values for the energy gap between $\Upsilon(1S)$ and unbound states that are bigger than the experimental value of the difference between the mass of $\Upsilon(1S)$ and two $B$ mesons. Therefore, we assume that at such small temperatures there are no sizeable thermal effects. 

\begin{figure}
\includegraphics[scale=0.5]{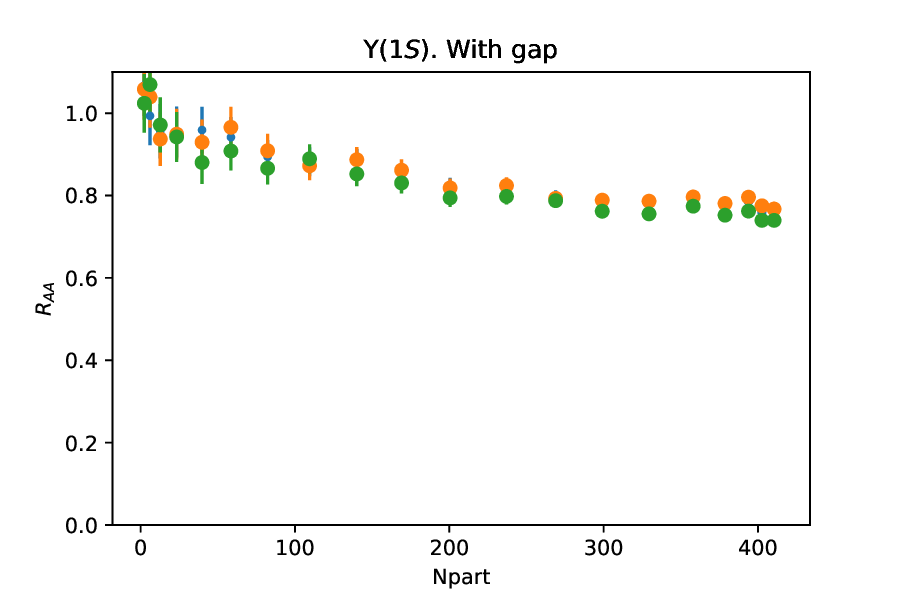}\includegraphics[scale=0.5]{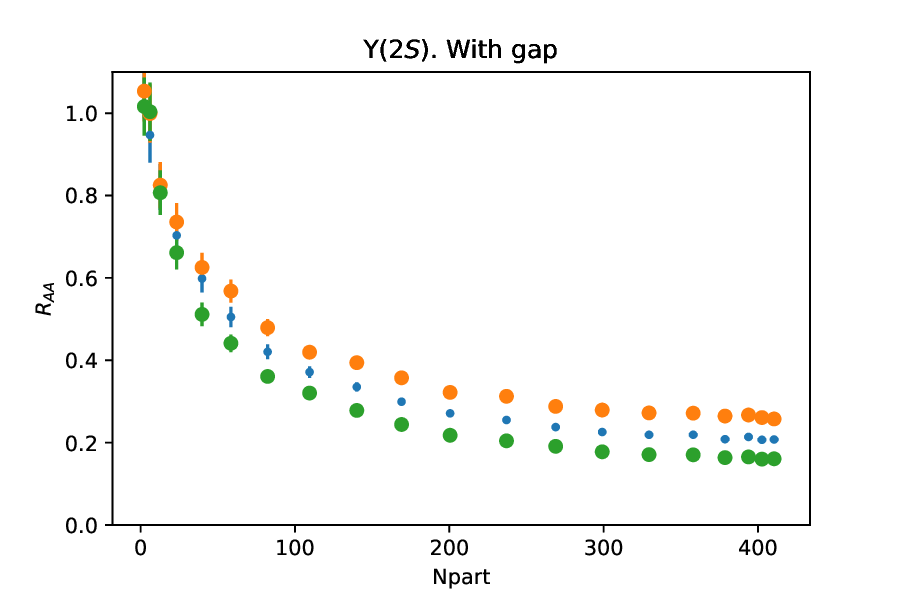}
\caption{Prediction for $R_{AA}$ of $\Upsilon(1S)$ (left) and $\Upsilon(2S)$ (right)  in the lattice  scenario. The different symbols correspond to different parameters shown in table \ref{tb:st}. The blue, orange and green points correspond respectively to the S1, S2 and S3 set of parameters. Physically, the three scenarios correspond to considering the central value of $m_D$ or the lower (larger) value of $m_D$ compatible with the error.}
\label{fig:st1raavsnpart_1s}
\end{figure}

\begin{figure}
\includegraphics[scale=0.5]{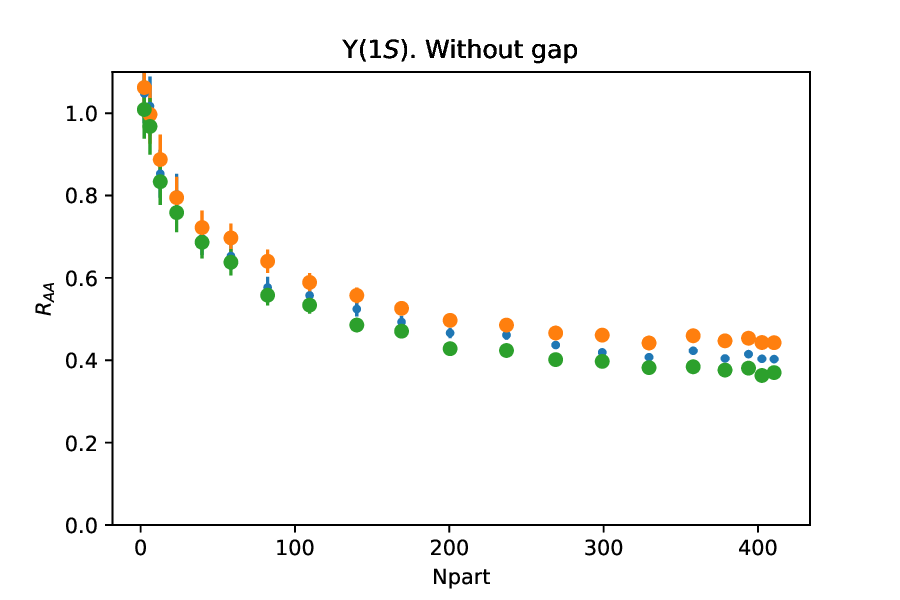}\includegraphics[scale=0.5]{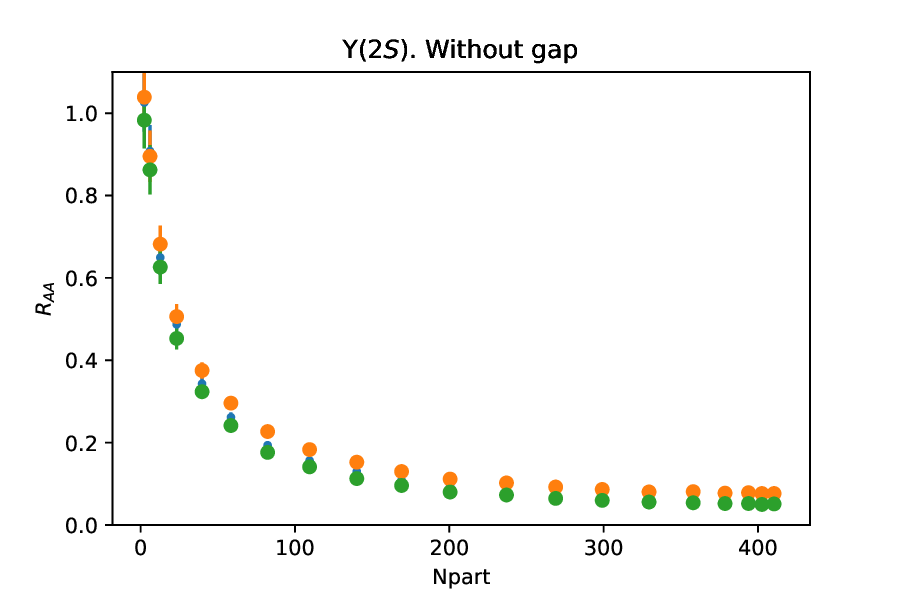}
\caption{Same as Figs. \ref{fig:st1raavsnpart_1s} but using $\Gamma_{\rm s}^0$ instead of $\Gamma_{\rm s}$.}
\label{fig:st1raavsnpart_1sgl}
\end{figure}

Knowing this, we obtain the $R_{AA}$ results shown in Fig.~\ref{fig:st1raavsnpart_1s}. In the case of $\Upsilon(1S)$, the condition that the binding energy is much bigger than the decay width is fulfilled for a wide range of centralities. However, for $\Upsilon(2S)$ this is not the case. In fact, the validity range of our approximations for $\Upsilon(2S)$ in the lattice scenario is similar to the one of $\Upsilon(1S)$ in the perturbative case. In Fig. \ref{fig:st1raavsnpart_1sgl} we do the same but ignoring the gap. As in the perturbative case, we see that the difference is quite substantial (again a factor $\sim 2$). Comparing Fig.~\ref{fig:st1raavsnpart_1sgl} and Fig.~\ref{fig:st1raavsnpart_1s} one can see that this factor can be essential to bring the $R_{AA}$ in agreement with its experimental value. This is approximately the case in Fig.~\ref{fig:st1raavsnpart_1s}, although no strong conclusion can be drawn from this remark, given the simplifications made in the present analysis. In fact, the results in Fig.~\ref{fig:st1raavsnpart_1s} slightly underpredict the experimentally observed suppression, but this includes, among other things, a sizeable contribution from cold nuclear matter effects that are ignored in the present analysis. 
\section{Conclusions}
\label{sec:conclu}
In this work, we have explored the phenomenological consequences of the observations made in \cite{Blaizot:2018oev}, where we  highlighted the importance of taking into account  the energy gap between singlet and octet states when computing the decay width of a quarkonium bound state in a medium. The role of this energy gap can be understood as an energy dependence in the imaginary potential used to determine the bound state properties. The discussion in \cite{Blaizot:2018oev} was limited to the case of a static medium.  In the present study, we have extended the analysis to the case of an expanding medium, and explored the predictions of a simple  model in two different scenarios, one based on perturbative formulae, and a scenario based on the  lattice QCD data of \cite{Lafferty:2019jpr}. Our results have corroborated our initial findings, and indicate that the role of the energy dependence of the imaginary potential is as important in the expanding case as it is in the static medium: the  value of $R_{AA}$  is significantly lower when the energy dependence is ignored. The effect is sizeable, the energy dependence reducing the expected suppression by typically a factor 2 for bottomonium at present LHC energies. 

\begin{acknowledgments}
This study was triggered by discussions initiated in the EMMI Rapid Reaction Task Force “Suppression and (re)generation of quarkonium in heavy-ion collisions at the LHC”. M.A.E. received financial support from Xunta de Galicia (Centro singular de investigación de Galicia accreditation 2019-2022), the European Union ERDF, the “María de Maeztu” Units of Excellence program MDM2016-0692, the Spanish Research State Agency and from the European Research Council project ERC-2018-ADG-835105 YoctoLHC.
\end{acknowledgments}
\bibliography{Saclayap}
\end{document}